\documentclass[prl,reprint,amsmath,amssymb,aps]{revtex4-1}
\pdfoutput=1
\usepackage{fullpage}
\usepackage{graphicx}
\usepackage{braket}
\usepackage{siunitx}
\usepackage{dcolumn}
\usepackage{bm}
\newcommand{\be}{\begin{equation}}
\newcommand{\ee}{\end{equation}}
\newcommand{\bea}{\begin{eqnarray}}
\newcommand{\eea}{\end{eqnarray}}

\renewcommand{\epsilon}{\varepsilon}
\newcommand{\reffig}[1]{Fig.~\ref{#1}}
\newcommand{\reffigs}[1]{Figs.~\ref{#1}}
\newcommand{\refeq}[1]{Eq.~(\ref{#1})}
\newcommand{\refeqs}[2]{Eqs.~(\ref{#1}) and (\ref{#2})}

\begin{document}

\title{
Creating Complex Optical Longitudinal Polarization Structures
}
\author{F. Maucher$^{1,2}$, S. Skupin$^{3,4}$, S. A. Gardiner$^{1}$, I. G. Hughes$^{1}$\\ \ }
\affiliation{
  $^1$Joint Quantum Centre (JQC) Durham-Newcastle, Department of Physics, Durham University, Durham DH1 3LE, United Kingdom.\\
  $^2$Department of Mathematical Sciences, Durham University, Durham DH1 3LE, United Kingdom.\\ 
  $^3$Univ.~Bordeaux -- CNRS -- CEA, Centre Lasers Intenses et Applications, UMR 5107, 33405 Talence, France; \\
  $^4$Univ Lyon, Universit\'e Claude Bernard Lyon 1, CNRS, Institut Lumi\`ere Mati\`ere, F-69622, Villeurbanne, France
}

\begin{abstract}
In this paper we show that it is possible to structure the longitudinal polarization component of light. We 
illustrate our approach by demonstrating linked and knotted longitudinal vortex lines acquired upon 
non-paraxially propagating a tightly focused sub-wavelength beam. Remaining degrees of freedom in the transverse polarization components 
can be exploited to generate customized topological vector beams. 
\end{abstract}
\maketitle

The concept of light being a transverse wave represents an approximation that is suitable 
if the angular spectrum is sufficiently narrow~\cite{Born:95}. However, many practical applications 
ranging from microscopy to data storage 
require tight focusing. 
Tight focusing implies a broad angular spectrum and the notion of light being 
transverse becomes inappropriate. Hence, the longitudinal polarization component can typically not be neglected~\cite{Richards:PRS:1959,Youngworth:OE:00}. 
To mention a few examples, a ``needle beam'' with particularly large longitudinal component was proposed in \cite{Chong:NatPhot:2008}, 
and radial transverse polarization 
permits the significant decrease of the focal spot size~\cite{Leuchs:OC:2000,Leuchs:PRL:2003} while the generated longitudinal 
component may even dominate the interaction with matter~\cite{Hnatovsky2011}. Last but not least, a M{\"o}bius strip in the polarization of light was realized in~\cite{Bauer:Science:2015}.

In addition, there is current substantial interest in ``structured light'', that is, generating customized light fields that suit 
specific needs in applications in a range of fields~\cite{Allen:PRA:1992,structured_light,Franke-Arnold:17,Maucher2:PRL:2016}.
Since the proposal of the Gerchberg-Saxton algorithm~\cite{Gerchberg:Optik:1972} 1972, 
advances in light shaping~\cite{Grier:Nature:2003,Whyte:NJP:2005,Shanblatt:OE:11} now permit the realization of complex light 
patterns in the transverse polarization plane, including light distributions the optical vortex lines of which form knots~\cite{Dennis:RoyProcA:2001,Leach:NJP:2005,Dennis:Nature:2010}. 
Knotted topological defect lines and their dynamics have been studied in diverse other settings, 
including for example classical fluid dynamics~\cite{Moffatt:JFM:1969,Moffatt:nature:1990,Irvine:nature:2013}, excitable media \cite{Paul:PRE:2003, Maucher:PRL:2016, Maucher:PRE:2017}, 
and nematic colloids \cite{Tkalec:science:2011,Martinez:NatMat:2014}.
To date, the approach has typically been to determine the longitudinal polarization component of the electric field 
from given transverse components, and attempts to target complex structures in the longitudinal component have not yet been pursued. 
The reason for this is twofold. On the one hand, the longitudinal component is not directly accessible by beam shaping techniques. On the other hand, non-paraxial beam configurations are required, and topological light is usually studied in paraxial approximation. 
It is therefore not immediately evident that the whole range of three-dimensional light configurations known for transverse components can be realized in the longitudinal component as well. 

In this paper, we will show that complex light-shaping of the longitudinal polarization 
component is indeed possible. To this end, we firstly identify non-paraxial light patterns that give rise to vortex lines that form knots or links.
Secondly, we invert the problem and derive how one must structure the 
transverse components of a tightly focused beam to give rise to a {\em given} complex pattern in the longitudinal component, and thus 
present the first example of non-transverse non-paraxial
knots. Finally, we demonstrate that remaining degrees of freedom in the transverse polarization components 
allow for simultaneous transverse shaping, which could be interesting for applications, e.g., inscribing vortex lines into Bose-Einstein Condensates.

\begin{figure}
  \includegraphics[width=\columnwidth]{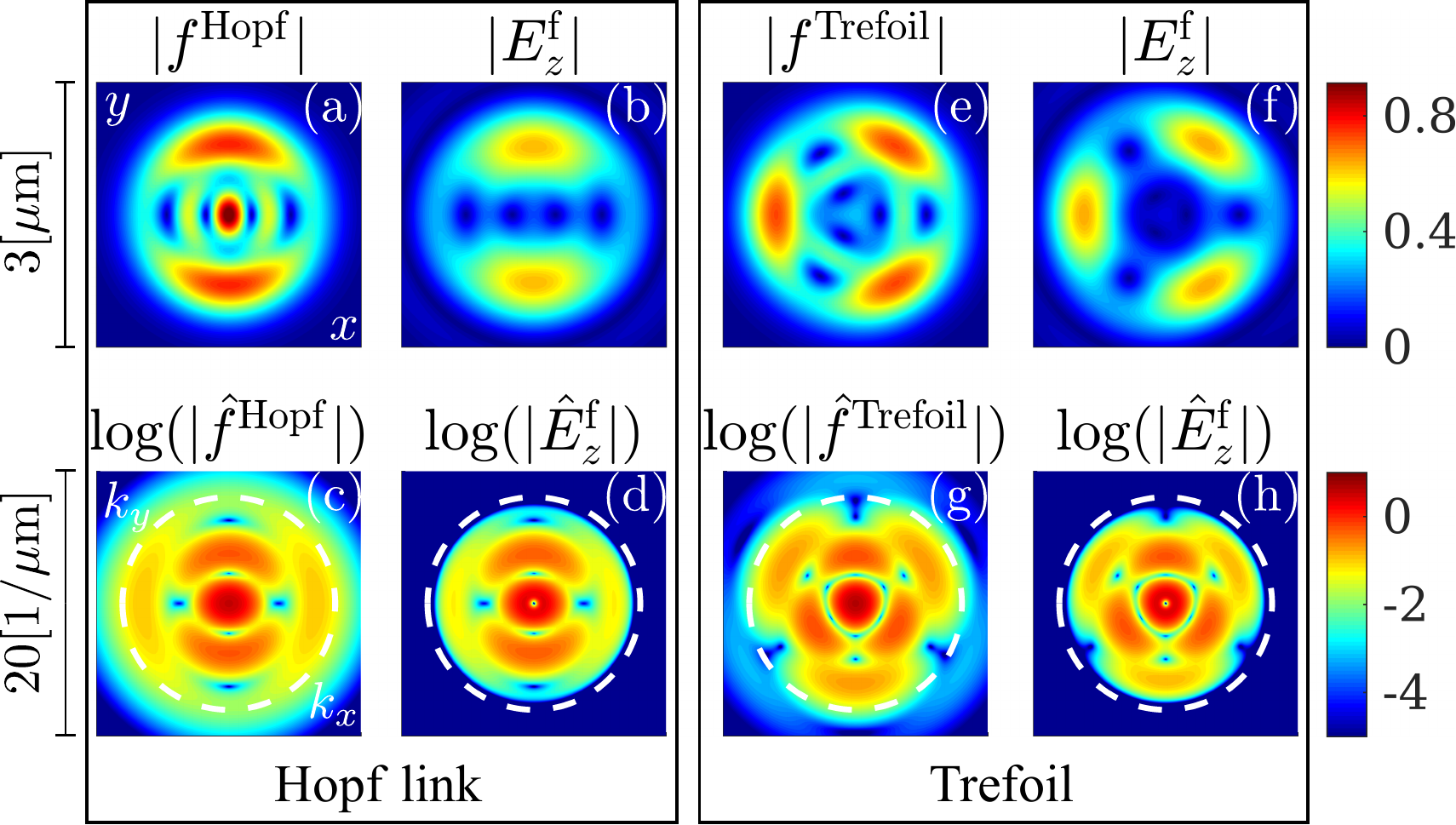}
 \caption{The profiles $f^{\mathrm{ Hopf}}$ and $f^{\mathrm{ Trefoil}}$ of \refeqs{eq:hopf}{eq:trefoil} (a,e) at narrow widths contain evanescent waves. This is demonstrated in (c,g), where the profiles are shown in the transverse Fourier domain together with a circle of radius $k_0$. A spectral attenuation (d,h) according to \refeq{eq:attenuation} removes the 
 evanescent amplitudes as well as amplitudes close to $\bm k_\perp=0$ (see text for details), and alters $E^{\rm f}_z$ in the focal plane significantly (b,f).}
\label{fig:hopf_trefoil_evanescent}
\end{figure}

We begin with the equations describing a monochromatic light beam:
\begin{align}
 \nabla^2{\bf E}({\bf r}_{\perp},z) + {k}^2_0{\bf E}({\bf r}_{\perp},z) &=0, \label{eq:nonparaxial}\\
 \nabla\cdot{\bf E}({\bf r}_{\perp},z) =\nabla_{\perp}\cdot{\bf E}_{\perp}+\partial_z E_z&= 0.\label{eq:divE}
\end{align}
Here, $k^2_0=\omega^2/c^2=(2\pi/\lambda)^2$, and we have introduced the transverse coordinates ${\bf r}_{\perp}=(x,y)$ and transverse electric field components ${\bf E}_{\perp}=(E_{x},E_{y})$  as we consider propagation in the positive $z$ direction.
All three components of ${\bf E}$ in~\refeq{eq:nonparaxial} fulfil the same wave equation, and for a given field configuration 
${\bf E^{\rm f}}({\bf r}_{\perp})$ at $z=0$ (e.g., at focus) the general solution for propagation in the positive $z$ direction reads $\hat{{\bf E}}({\bf k}_{\perp},z)=\hat{\bf {E}}^{\rm f}({\bf k}_{\perp})\exp(i k_z z)$, where $k_z({\bf k}_{\perp})=\sqrt{k_0^2-{\bf k}_{\perp}^2}$, ${\bf k}_{\perp} = (k_x,k_y)$ and the symbol $\hat{~}$ denotes the transverse Fourier domain. 
The prescribed field configuration ${ \hat{\bf E}^{\rm f}}$ must obey certain constraints. Firstly, in order to get a valid bulk solution, 
there must be no evanescent fields present, that is, ${ \hat{\bf E}^{\rm f}}=0$ for ${\bf k}_{\perp}^2\ge k_0^2$. Secondly, Eq.~(\ref{eq:divE}) implies for solutions propagating in the $z$ direction that ${\hat{E}_z^{\rm f}}({\bf k}_{\perp}=0)=0$.

As preparation for what follows, we first 
investigate how to obtain a non-paraxial tightly focused knot or link in $E_z$, assuming that we can directly prescribe ${E}_z^{\rm f}$.
For the transverse paraxial case, recipes to generate vortex lines in various shapes are known, and they usually involve 
linear combinations of Laguerre-Gaussian modes~\cite{Dennis:Nature:2010,Maucher:NJP:2016}. 
These recipes are not directly applicable to our problem, since there are evanescent fields, due to the nonparaxiality the wavelength cannot be scaled away and it would lead to Eq.~(\ref{eq:divE}) being violated. 
Nevertheless, we found that it is possible to adopt those recipes for the non-paraxial case by an educated guess.
Starting from a given linear combination of Laguerre-Gaussian modes $f$, filtering in the transverse Fourier domain~\cite{Goodman:16},
$$
H_{k_0}(\bm k_\perp)=e^{-\frac{1}{ 2\lambda^2\left(\sqrt{\bm k_\perp^2}-k_0\right)^2}}, 
H_{0}(\bm k_\perp)=1-e^{-(3\lambda{\bm k_\perp})^2},
$$
chops off evanescent amplitudes as well as amplitudes close to $\bm k_\perp=0$, and the longitudinal polarization component at $z=0$ reads 
\begin{equation}
 \hat E_z^{\rm f}(\bm k_\perp) = 
 \begin{cases}
 \hat f(\bm k_\perp) H_{k_0}H_{0},& \text{for } \bm k_\perp^2<k_0^2 \\
  0 &\text{for } \bm k_\perp^2\geq k_0^2
 \end{cases}.
   \label{eq:attenuation}
\end{equation}
Since the higher-order Laguerre-Gaussian modes are broader in Fourier space and thus lose relative weight after attenuation,
one must decrease the relative amplitudes of the lower-order modes to a certain extend.
We have found that the following field structures produce a Hopf link or trefoil, respectively, 
\begin{align}
 f^{\mathrm{ Hopf}}&=4{\rm LG}_{00}^\sigma-5{\rm LG}_{01}^\sigma +11{\rm LG}_{02}^\sigma-8{\rm LG}_{20}^\sigma
 \label{eq:hopf}\\
 \begin{split}
 f^{\mathrm{ Trefoil }}&=9{\rm LG}_{00}^\sigma- 20 {\rm LG}_{01}^\sigma + 40{\rm LG}_{02}^\sigma \\
 & \quad- 18{\rm LG}_{03}^\sigma - 34 {\rm LG}_{30}^\sigma, 
 \end{split}\label{eq:trefoil}
\end{align}
for wavelength $\lambda=780~{\rm nm}$ and width  $\sigma=370~{\rm nm}\approx \lambda/2$ of the usual Laguerre-Gaussian modes ${\rm LG}^{\sigma}_{ij}(\bm r_\perp)$. 
The resulting amplitudes before and after filtering for $f$ being either $f^{\mathrm{ Hopf}}$ or $f^{\mathrm{ Trefoil }}$
defined in \refeqs{eq:hopf}{eq:trefoil} are plotted in \reffig{fig:hopf_trefoil_evanescent} after normalization to unity.
We note that individual mode amplitudes can be 
changed by about $10\%$ without altering topology, 
demonstrating a degree of robustness and hence experimental feasibility.

\begin{figure}
 \includegraphics[width=\columnwidth]{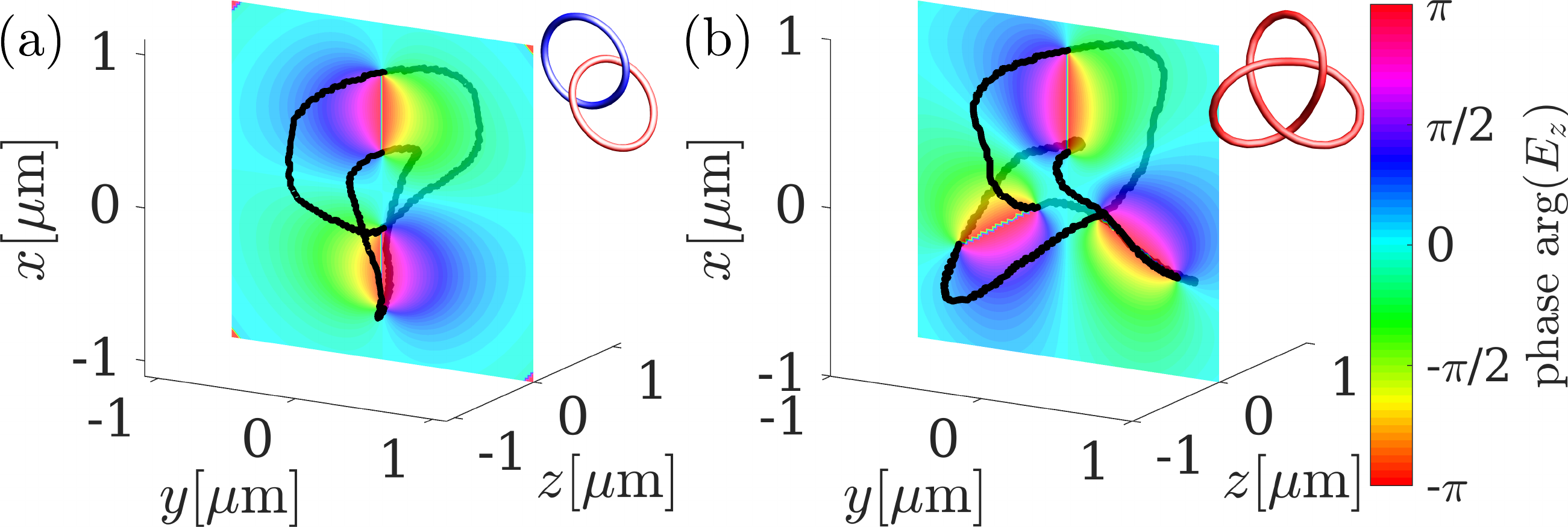}
 \caption{Propagation of the spectrally attenuated field shown in \reffigs{fig:hopf_trefoil_evanescent}(b,f) gives rise
 to the vortex lines (black) in the forms of a Hopf link (a) and a trefoil (b). A phase slice is shown in the $xy$ plane at $z=0$. 
 As comparison, idealised Hopf link and trefoil are shown as insets.
 }
\label{fig:hopf_trefoil_propagation}
\end{figure}

Let us now verify that the presented patterns in the focal plane in fact give rise to vortex lines with the desired topology. 
It is straightforward to propagate the filtered component $E^{\rm f}_z$, as defined in \refeq{eq:attenuation}, in the $z$-direction.
The vortex lines throughout three-dimensional space  are depicted by the black lines in 
\reffigs{fig:hopf_trefoil_propagation}(a,b) together with a slice in the $z=0$ plane of the light profile phase.
The obtained vortex lines are topologically equivalent to a Hopf link and a trefoil, as drawn in the insets.

\begin{figure}
 \includegraphics[width=\columnwidth]{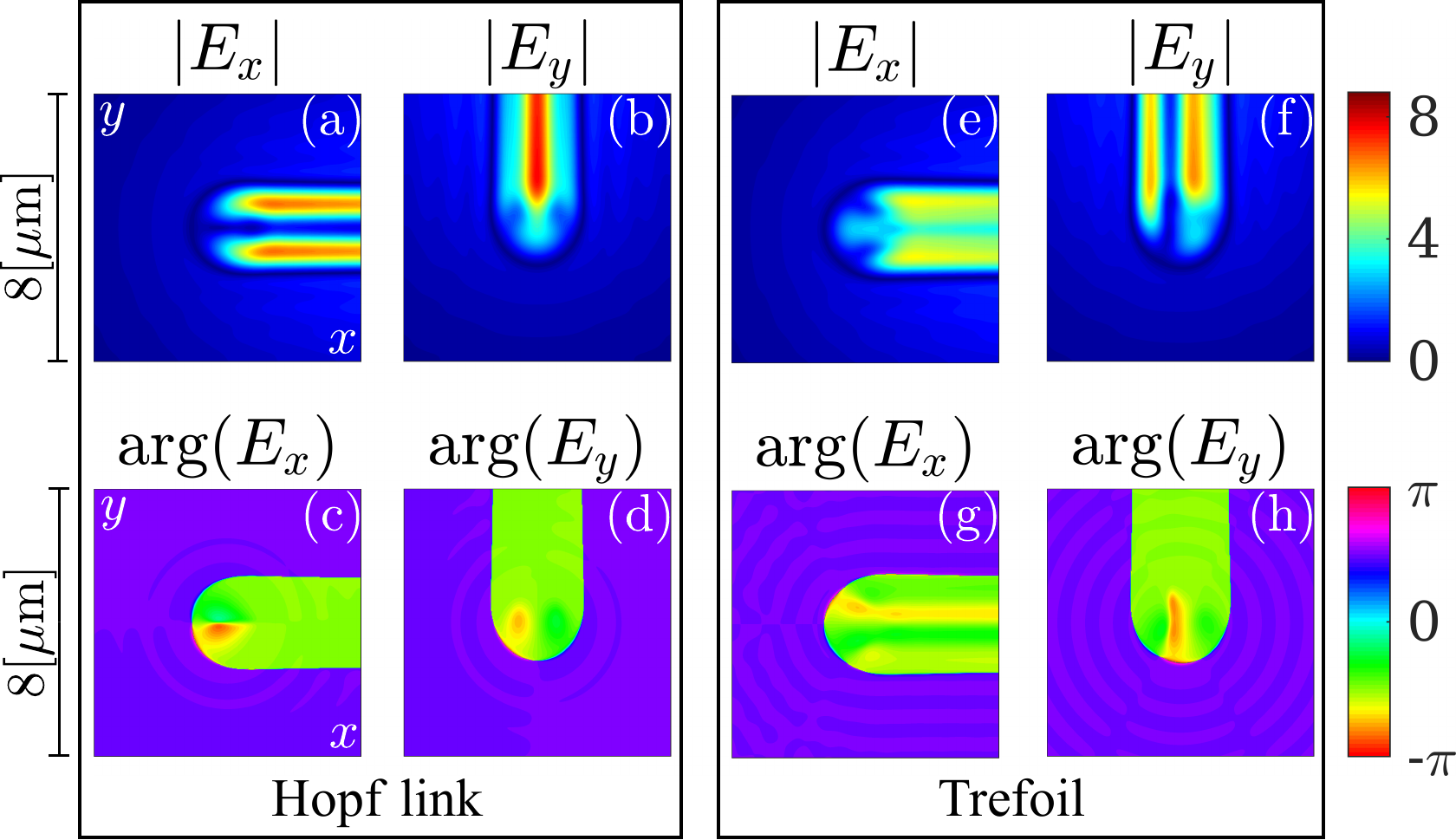}
 \caption{
 Amplitude and phase for the linearly polarized transverse components \refeqs{eq:xpol}{eq:ypol} that give rise to a longitudinal component forming the Hopf link~\reffig{fig:hopf_trefoil_evanescent}(b),~\reffig{fig:hopf_trefoil_propagation}(a) is
 shown in (a--d) and the trefoil~\reffig{fig:hopf_trefoil_evanescent}(f),~\reffig{fig:hopf_trefoil_propagation}(b) is shown in (e--h). 
 The colormaps for each figure are on the right of the two rows of plots. 
 }
\label{fig:hopf_trefoil_focus}
\end{figure}

We now address the main point of this paper, i.e. how to choose the transverse polarization components to obtain a given longitudinal polarization component.
Because only the transverse components $E_x$ and $E_y$ are accessible to beam shaping, this point is also of great practical relevance.
When inspecting \refeq{eq:divE}, at a first glance the problem may seem to be ill-posed, given that only the longitudinal derivative of the 
longitudinal polarization enters, i.e., $\partial_zE_z$. 
However, in Fourier space it is easy to see from \refeq{eq:divE} that a linearly polarized solution to this problem is given by 
\begin{equation}
 E_x^{\rm f}  =-i\!\int\limits_{-\infty}^x\! \mathcal{F}^{-1} \left[ k_z \hat{E}_z^{\rm f} \right]\!(x^\prime,y)dx^\prime, \quad E_y^{\rm f}  = 0,
 \label{eq:xpol}
\end{equation}
where $\mathcal{F}^{-1}[\hat{g}](x,y)=g(x,y)$ denotes the inverse transverse Fourier transformation. Obviously, 
an orthogonally polarized solution also exists,
\begin{equation}
 E_x^{\rm f}  = 0, \quad E_y^{\rm f} = -i\!\int\limits_{-\infty}^y\! \mathcal{F}^{-1} \left[ k_z \hat{E}_z^{\rm f} \right]\!(x,y^\prime)dy^\prime.
 \label{eq:ypol}
\end{equation}
Both $x$  and $y$ polarized solutions \refeqs{eq:xpol}{eq:ypol} evaluated for Hopf link and trefoil are depicted in Fig.~\ref{fig:hopf_trefoil_focus}.
It is noteworthy that any superposition of real and imaginary parts of the solutions~\refeqs{eq:xpol}{eq:ypol} is admissible, as long as the 
coefficients of this superposition add up to one. Furthermore, an arbitrary solenoidal field can be added without having an effect on the longitudinal 
component. We will discuss this later in more detail.

Unfortunately, the transverse polarization components computed from \refeqs{eq:xpol}{eq:ypol} are impractical, since, even though
$E^{\rm f}_z$ has finite support, 
the components $E^{\rm f}_x$ or $E^{\rm f}_y$ are non-zero on a semi-infinite interval (see Fig.~\ref{fig:hopf_trefoil_focus}). 
However, simply attenuating these components by multiplying with, e.g., a sufficiently wide super Gaussian profile
${\rm SG}_N^{w}({\bm r}_{\perp})=\exp(-{\bm r}_{\perp}^{2N}/w^{2N})$ allows the resolution of the problem of semi-infinite light distributions without affecting 
propagation of the longitudinal component close to the optical axis. Evaluating $\nabla_{\perp}\cdot\left[{\rm SG}_N^{w}({\bm r}_{\perp}){\bf E}^{\rm f}_{\perp}({\bm r}_{\perp})\right]$ reveals that, where $\nabla_{\perp}{\rm SG}_N^{w}$ is large and points in the direction of ${\bf E}^{\rm f}_{\perp}$,
additional satellite spots in the longitudinal component will appear. We have checked that using e.g. a super Gaussian with $N=5$ and $w=10 \lambda$ ensures that these additional spots are sufficiently far from the region of interest and both Hopf link and trefoil develop in the propagation of the modified $E_z$ component. 

\begin{figure}
\includegraphics[width=\columnwidth]{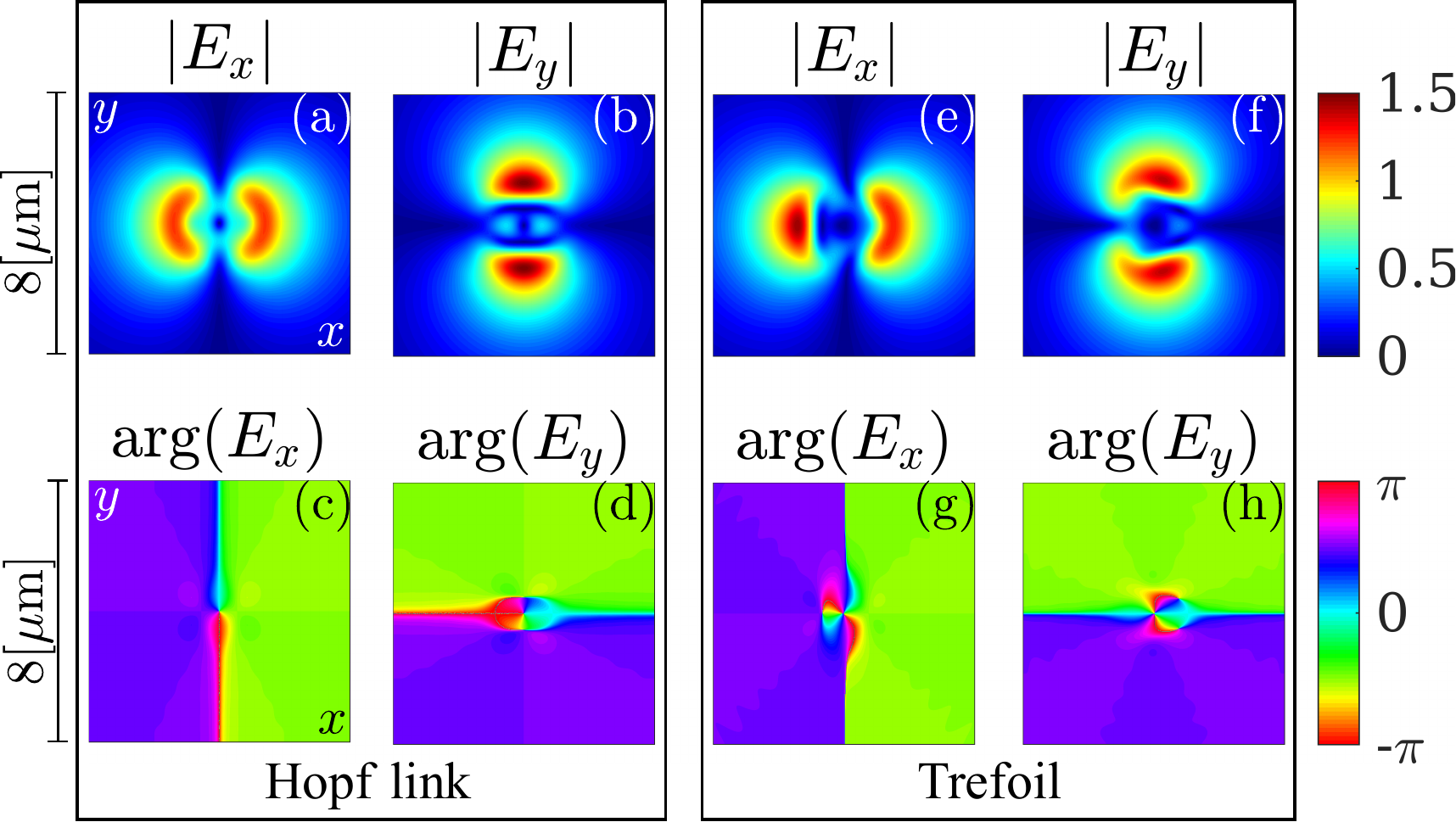}
 \caption{Irrotational transverse amplitude and phase profiles producing a Hopf link (a--d) and a trefoil (e--h) in the longitudinal polarization component. 
 }
\label{fig:optimized}
\end{figure}

So far, we have seen that the answer to the problem of how to choose ${\bf E}^{\rm f}_{\perp}(\bm r_\perp)$ for realizing a prescribed $E^{\rm f}_z$ is not unique, and there are certain degrees of freedom in the choice of ${\bf E}^{\rm f}_{\perp}$. 
The fundamental theorem of vector calculus (Helmholtz decomposition) allows us to decompose a (sufficiently well-behaved) vector field ${\bf F}$ into an 
irrotational (curl-free) and a solenoidal (divergence-free) vector field, 
and ${\bf F}$ can be written as ${\bf F}=-\nabla \phi+\nabla\times{\bf A}$. We wish to apply this theorem to the transverse plane, that is, we set ${\bf F}={\bf E}^{\rm f}_{\perp}(\bm r_\perp)$. In this case, the decomposition reduces to 
\begin{equation}
{\bf E}^{\rm f}_{\perp}(\bm r_\perp) = -\nabla \phi(\bm r_\perp) + \nabla \times \left[A(\bm r_\perp){\bf e}_z\right], \label{eq:helmholtz}
\end{equation}
with ${\bf e}_z$ the unit vector in $z$ direction. It is straightforward to verify that 
\begin{equation}
 \hat \phi({\bf k}_{\perp}) = - i\frac{k_z({\bf k}_{\perp})\hat{ E}_z^{\rm f}({\bf k}_{\perp})}{{\bf k}_{\perp}^2} \label{eq:phi}
\end{equation}
gives rise to a valid transverse polarization component ${\bf E}^{\rm f}_{\perp}$. The scalar function $A(\bm r_\perp)$ may be chosen arbitrarily,
because the term ${\bf E}^{\rm sol}_{\perp} = \nabla \times \left[A(\bm r_\perp){\bf e}_z\right]$ produces the solenoidal part of ${\bf E}^{\rm f}_{\perp}$, which does not give
rise to any longitudinal polarization component.
The irrotational choice for ${\bf E}^{\rm f}_{\perp}$, that is, evaluating \refeqs{eq:helmholtz}{eq:phi} with $A(\bm r_\perp)=0$,
for Hopf link and trefoil are shown in \reffig{fig:optimized}.

While the transverse polarization components shown in Figs.~\ref{fig:hopf_trefoil_focus} and \ref{fig:optimized} produce exactly the same 
longitudinal field, they are completely different, in particular from a topological point of view.
Unlike the light distributions in~\reffig{fig:hopf_trefoil_focus}, which do not contain vortices in the transverse 
polarization components, the irrotational transverse polarization components in~\reffig{fig:optimized} each feature
phase singularities. 
Furthermore, note that the amplitudes required in the irrotational transverse polarizations are only roughly two to three times 
the peak amplitude in the longitudinal polarization. 

We have seen that all possible transverse polarization components producing a certain longitudinal component differ 
by a solenoidal field ${\bf E}^{\rm sol}_{\perp} = \nabla \times \left[A(\bm r_\perp){\bf e}_z\right]$, 
and the function $A(\bm r_\perp)$ represents the degrees of freedom one has when shaping the ${\bf E}^{\rm f}_{\perp}$. 
As our examples show, it is possible to control the topological structure of longitudinal and transverse electric field components simultaneously. 
Tightly focused beams containing vortex lines play a role in inscribing vortex lines with specific 
topology into Bose-Einstein condensates using two-photon Rabi-transitions~\cite{Ruostekoski:PRA:2005,Maucher:NJP:2016}. 
The demonstrated knotted or linked longitudinal vortex lines have an extent of roughly $1~\mu$m$^3$ and thus match the typical 
size of a Bose-Einstein Condensate. Being able to exploit the unique features of structured light in all three vector components of the electric field 
opens new avenues in controlling the interaction of light with matter.

An important practical issue is to actually experimentally detect such a small structure in the longitudinal polarization component. 
Probing of the longitudinal field using molecules was achieved experimentally roughly 15 years ago~\cite{Novotny:PRL:2001} 
and continues to be of interest for light-matter interactions~\cite{Quinteiro:PRL:2017}.
We propose using a tomographic method using a thermal Rubidium vapour cell that is very thin compared to the wavelength~\cite{Sargsyan:OL:17} to
experimentally access the longitudinal polarization component. 
Using an additional strong static magnetic field parallel to the optical axis and tuning the light field to resonantly drive a $\pi$-transition 
allows the selective coupling of the longitudinal polarization only. 
To separate the $\pi$-transition from the $\sigma^{\pm}$-transitions beyond Doppler broadening 
(roughly 0.5GHz at \SI{100}{\celsius}) we need a sufficiently large magnetic field (roughly $B\sim 1T$). 
For such large magnetic fields, isolated pure $\pi$-transitions exist 
e.g.\ from $\ket{{\mathrm 5S_{1/2}m_jm_I}}=\ket{{\mathrm 5S_{1/2}\pm1/2\pm3/2}}$ to 
$\ket{{\mathrm 5P_{3/2}\pm1/2\pm3/2}}$. 
This method of light-matter coupling can however be extended to more general settings, where the angle of the magnetic field can be 
tuned and thus different components of vectorial topological light can superposed and inscribed into matter. 

In conclusion, we have presented a simple algorithm to realize an arbitrary (sufficiently well-behaved) field in the focal plane in the longitudinal 
polarization component, and elaborated on how to realize the transverse components for it. 
We have highlighted the importance of the occurrence of evanescent waves and discussed the important degrees of freedom 
in the choice of the transverse polarization components.  
Using this method has the potential to broaden the range of possible vectorial structured light fields extensively and lead to 
a range of applications in various fields in physics, including nonlinear optics and Bose-Einstein Condensates. 

\acknowledgments 
This work is funded by the Leverhulme Trust Research Programme Grant RP2013-K-009, SPOCK: Scientific Properties Of Complex Knots.
S.S.\ acknowledges support by the Qatar National Research Fund through the National Priorities Research Program (Grant No.\ NPRP 8-246-1-060).

\bibliographystyle{apsrev4-1}
\bibliography{bib}

\end{document}